# Electric Contributions to Magnetic Force Microscopy Response from Graphene and MoS$_2$ Nanosheets

Lu Hua Li[1]* and Ying Chen[1]

[1]Institute for Frontier Materials, Deakin University, Geelong Waurn Ponds Campus, Victoria 3216, Australia

*Corresponding author: luhua.li@deakin.edu.au

ABSTRACT

Magnetic force microscopy (MFM) signals have recently been detected from whole pieces of mechanically exfoliated graphene and molybdenum disulfide (MoS$_2$) nanosheets and magnetism of the two nanomaterials was claimed based on these observations. However, non-magnetic interactions or artefacts are commonly associated with MFM signals, which makes the interpretation of MFM signals not straightforward. A systematic investigation has been done to examine possible sources of the MFM signals from graphene and MoS$_2$ nanosheets and whether the MFM signals can be correlated with magnetism. It is found that the MFM signals have significant non-magnetic contributions due to capacitive and electrostatic interactions between the nanosheets and conductive cantilever tip, as demonstrated by electric force microscopy (EFM) and scanning Kevin probe microscopy (SKPM) analyses. In addition, the MFM signals of graphene and MoS$_2$ nanosheets are not responsive to reversed magnetic field of the magnetic cantilever tip. Therefore, the observed

MFM response is mainly from electric artefacts and not compelling enough to correlate with magnetism of graphene and $MoS_2$ nanosheets.

## I. INTRODUCTION

The presence of permanent magnetism in graphite (including graphene) and other carbon-based materials is towards consensus after decades of debate. Weak ferromagnetism has been detected from both pristine and irradiated highly-oriented pyrolytic graphite (HOPG) using magnetic force microscopy (MFM) and superconducting quantum interference device (SQUID).[1-5] The magnetism in graphite originates from point defects and edge states, according to theoretical and experimental investigations. Magnetism in graphene is also of great interest considering its promising use in spintronics and magnetoelectronics. Ferromagnetism has been detected from graphene because of defect and edge state,[6,7] strain,[8,9] doping,[10,11] functionalization[12,13] and substrate interaction,[14] though some studies found pristine graphene non-magnetic.[15] Other nanomaterials, such as molybdenum disulfide ($MoS_2$) nanosheets, could also gain magnetism due to similar reasons.[16-20]

Not until very recently has MFM been used to investigate mechanically exfoliated graphene and $MoS_2$ nanosheets for the first time.[21] MFM phase and amplitude contrasts have been detected from whole pieces of graphene and $MoS_2$ nanosheets of thickness up to 183 nm and vary depending on the nanosheet thickness. The observed MFM signals were used to claim magnetism of graphene and $MoS_2$ nanosheets. Unfortunately, no necessary investigation has been performed to examine possible non-magnetic contributions or artefacts to the MFM signals. In fact, MFM signals are often associated with non-magnetic contributions or artefacts, such as electric interactions, which sometimes could be predominant.[22] Therefore, to correctly interpret MFM signals, possible non-magnetic interactions must be carefully investigated. Here, the source of the reported MFM signals from graphene and MoS$_2$ nanosheets has been systematically investigated and the correlation between the MFM signals and magnetism is discussed.

**II. EXPERIMENTAL**

The graphene and MoS$_2$ nanosheets used in this work were exfoliated from HOPG (Ted Pella) and MoS$_2$ particles (Sigma-Aldrich) using the Scotch tape method. An Olympus BX51 optical microscope was used for the visualization of the few-layer graphene and MoS$_2$ nanosheets. A Cypher scanning probe microscope (Asylum Research) was used for the MFM, EFM and SKPM measurements. A Co/Cr coated magnetic cantilever with a spring constant of 2 N/m and coercivity of ~400 Oe (ASYMFM, Asylum Research) and Pt/Ti coated conductive cantilevers with a spring constant of 2 N/m (AC240TM, Asylum Research) were used. The magnetic cantilever tip was magnetized and also reversely magnetized by approaching it to a neodymium magnet either face-up or face-down.

**III. RESULTS AND DISCUSSION**

The preparation procedure of graphene and MoS$_2$ nanosheets and experimental conditions for MFM measurements are similar to those reported in Ref.[21]. The graphene and MoS$_2$ nanosheets were prepared on silicon wafer covered with a 90 nm layer of silicon oxide (90 nm SiO$_2$/Si) by the Scotch tape exfoliation and then visualized by optical microscope observation.[23-25] Figure 1a shows the optical microscopy photo of the exfoliated graphene nanosheets of different thicknesses. The AFM image in Figure 1b shows that they are monolayer (1L, height = 0.45 nm), bilayer (2L, 0.66 nm), trilayer (3L, 1.06 nm), 4-layer (4L, 1.36 nm) and 10-layer (10L, 3.35 nm) nanosheets (Figure 1b). The MFM measurements were conducted using a lift mode, in which the Co/Cr coated magnetic cantilever oscillating at its resonant frequency first scanned over the sample surface for topography and then was raised up for a constant height above the sample surface to detect long-range magnetic interactions between the magnetic cantilever tip and the sample surface by monitoring the phase change of the cantilever.

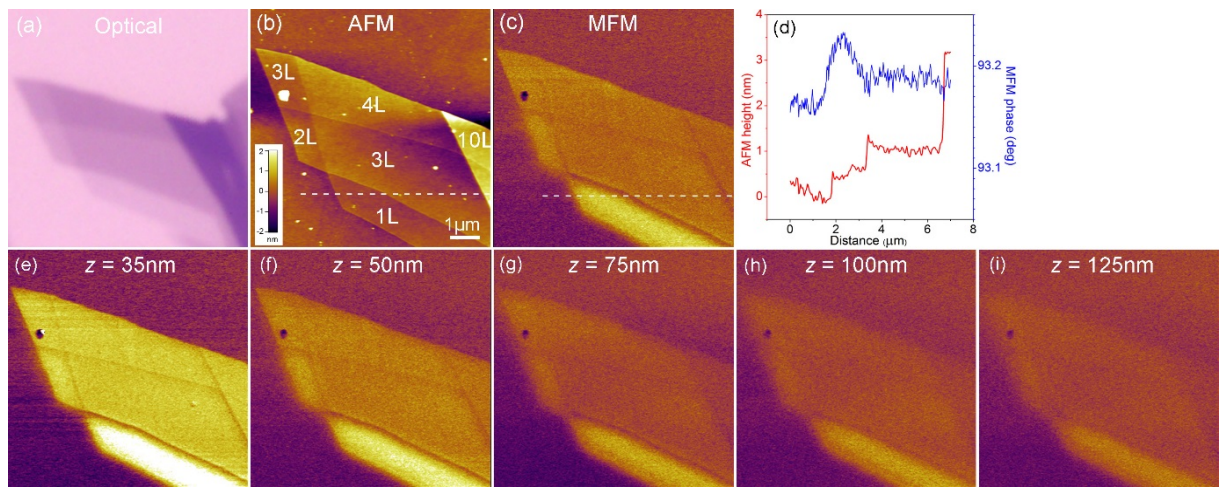

Figure 1. (a) Optical microscopy photo of the graphene nanosheets on 90nm $SiO_2$/Si substrate; (b) AFM topography image of the graphene sample which contains 1L, 2L, 3L, 4L and 10L nanosheets; (c) MFM phase image of the graphene nanosheets using a 50 nm lift height; (d) AFM height trace (red) and MFM phase shift trace (blue) from the dash lines in (b) and (c); (e-i) MFM phase images of the graphene nanosheets at different lift heights ($z$ = 35 to 125 nm). The scale range for all MFM phase images is 0.15°.

The obtained MFM results are consistent with those in the previous report.[21] As shown in Figure 1c, the graphene nanosheets do show MFM phase shift/contrast compared to the non-magnetic $SiO_2$/Si substrate, especially the 1L graphene showing a relatively strong contrast (note that positive MFM phase shift is defined as "attractive" in Cypher AFM, which is opposite to the setting in most other AFMs, *e.g.* Bruker AFM). In addition, similar reduction of the MFM phase shift with the increase of the lift height was observed in our experiment (Figure 1d-h), which was also reported in Ref.[21].

To test whether the observed MFM phase contrast is due to magnetic interactions or non-magnetic artifacts, the following three additional tests were conducted. MFM signals are often associated with non-magnetic interactions, especially long-range electric forces such as electrostatic and capacitive forces, two of the strongest non-magnetic contributions in MFM.[22] These two electric interactions could be prevalent for all surfaces no matter that it is conducting, semiconducting or insulating. So the first test is to evaluate the electrostatic and capacitive contribution to the MFM phase shift of the graphene. Capacitive and electrostatic contributions to MFM signals can be analyzed by applying DC voltages to the cantilever in the range of several volts. If magnetic interactions are prevailing, MFM images should not change dramatically under the influence of such small DC voltages, and vice versa.[22] Even when small DC voltages (–0.5 or +0.5 V) were applied to the magnetic cantilever during MFM measurements (the magnetic cantilever is conductive due to the Co/Cr coating), the MFM phase of the graphene was greatly affected (Figure 2): the 1L graphene had larger phase shift than the 10L graphene nanosheets at 0 V (Figure 2d), but the phase shift of the 1L graphene became smaller at –0.5 V. These results strongly suggest that capacitive and electrostatic interactions play important roles in the observed MFM phase response.

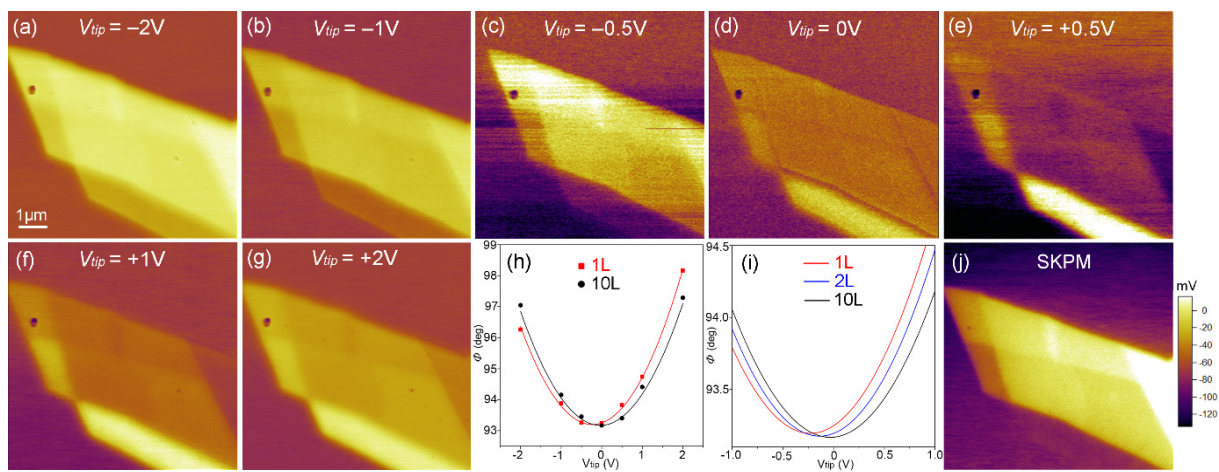

Figure 2. (a-g) MFM phase images of the graphene nanosheets under different tip voltages ($V_{tip}$ = –2 to +2 V) and tip lift height is 50 nm. The scale ranges: 2.5° for (a) and (g), 0.8° for

(b) and (f), 0.2° for (c-e); (h) second degree polynomial fittings of the MFM phase values from the 1L and 10L graphene nanosheets when $V_{tip}$ sweeping from –2 to +2 V; (i) fitted MFM phase of the 1L, 2L and 10L graphene nanosheets as a function of $V_{tip}$ to show their vertexes of parabolas located at $V_{tip} \neq 0$V. (j) SKPM image of the graphene nanosheets using the same tip lift height.

Fitting of the MFM phase shifts with respect to DC voltages can be used to estimate how significant the electric contributions are to the MFM images. If the cantilever frequency shift is smaller than its resonant frequency, the phase shift $\Delta\phi$ can be written as:[26]

$$\Delta\phi = \frac{\partial F/\partial z}{k} \cdot Q_{cant} \quad (1)$$

Where $\partial F/\partial z$ is the local force gradient or derivative of the force felt by the cantilever tip at a lift height $z$; $k$ is the spring constant of the cantilever; and $Q_{cant}$ is the quality factor or Q factor of the cantilever. In the case of capacitive and electrostatic interactions, if the capacitive interaction between the cantilever tip and the sample is treated as an ideal capacitance and the sample surface is planar, the force gradient can be simplified as:[26,27]

$$\partial F/\partial z = \frac{1}{2}\frac{\partial^2 C}{dz^2}\left(V_{tip} - V_{WFD} - V_Q\right)^2 - \frac{\partial C}{\partial z}E_Q V_{tip} \quad (2)$$

The first term relates to capacitive force, where $C$ is the local capacitance between the cantilever tip and the sample surface; $V_{tip}$ is the DC voltage applied to the cantilever tip; $V_{WFD}$ is the work function difference between the cantilever tip and the sample surface; $V_Q$ is the effective surface potential proportional to the trapped charges on the sample surface. The second term associates with Coulombic (electrostatic) force due to charges, where $E_Q$ is the charge-induced electric field applied to the cantilever tip. It can be seen that the capacitive part has a quadratic function to $V_{tip}$ and the Coulombic part is linear to $V_{tip}$. Because the Coulombic force is normally much smaller than the capacitive force, the force gradient and thus phase shift ($\Delta\phi$) due to capacitive and electrostatic forces generally show a parabola under varying $V_{tip}$.[28-30] Figure 2h shows that the measured phase shift values of the 1L and 10L graphene nanosheets can be well fitted using second degree polynomials, when $V_{tip}$ changes from –2 to +2 V. This indicates a good match to the model of capacitive and electrostatic interactions (Eq. 2). Therefore, electric interactions are predominant, if not all, in the observed MFM phase images.

The fittings of the phase shifts (Figure 2h) also show that the force gradients ($\partial F/\partial z$) of the capacitive and electrostatic interactions to the graphene nanosheets of different thicknesses are slightly different. For example, the vertexes of the parabolas (*i.e.* phase shift minima) are not at $V_{tip} = 0$ and different for the graphene nanosheets of different thicknesses. To view this more closely, Figure 2i compares the fitted parabolas of the phase shifts among 1L, 2L and 10L graphene nanosheets at around $V_{tip} = 0$. It can be seen that the vertex of the parabolic phase shift decreases with the graphene thickness: the vertex of the 1L graphene (red) has the most negative $V_{tip}$ value (–0.22 V) and the 10L graphene nanosheet (black) has the least negative value (–0.03 V). Such phenomenon can be better understood by studying the origins of the electrostatic and capacitive interactions, which are discussed below.

The electrostatic interactions from graphene nanosheets of different thicknesses can be explored using scanning Kevin probe microscopy (SKPM). The SKPM image in Figure 2j reveals that the graphene nanosheets of different thicknesses have different contrasts. The SKPM contrast could have two causes: (1) work function difference ($V_{WFD}$) among graphene nanosheets of different thicknesses and (2) different charge interactions between the cantilever tip and the graphene nanosheets. Because the $V_{WFD}$ between the cantilever tip and graphene with different thicknesses are quite similar: 4.57 eV for monolayer graphene, 4.69 eV for bilayer graphene and ~4.6 eV for bulk graphite,[31] the observed SKPM contrast should be mainly due to different charges on the surface of the graphene nanosheets of different thicknesses. The surface charges are due to unscreened charges from contaminations, e.g. dipolar water film, between the SiO$_2$ substrate and the graphene nanosheets or charge transfers between the substrate and the graphene.[29,32,33]

The capacitive interaction relates to work function difference ($V_{WFD}$), surface potential ($V_Q$) and local capacitance ($C$) (see Eq. 2). As discussed above, the work function difference among graphene nanosheets of different thicknesses is small and negligible, but the surface potentials due to trapped charges are quite different. In addition, it has been found that graphene nanosheets of different thicknesses have different capacitance.[34] So the capacitive contribution to the different phase shifts (*i.e.* force gradients) among graphene nanosheets of different thickness comes from their different surface potential ($V_Q$) and capacitance ($C$). These are the origins of the different phase contrasts for graphene nanosheets of different thicknesses and also explain why MFM phase contrast can be observed from the graphene (Figure 1c).

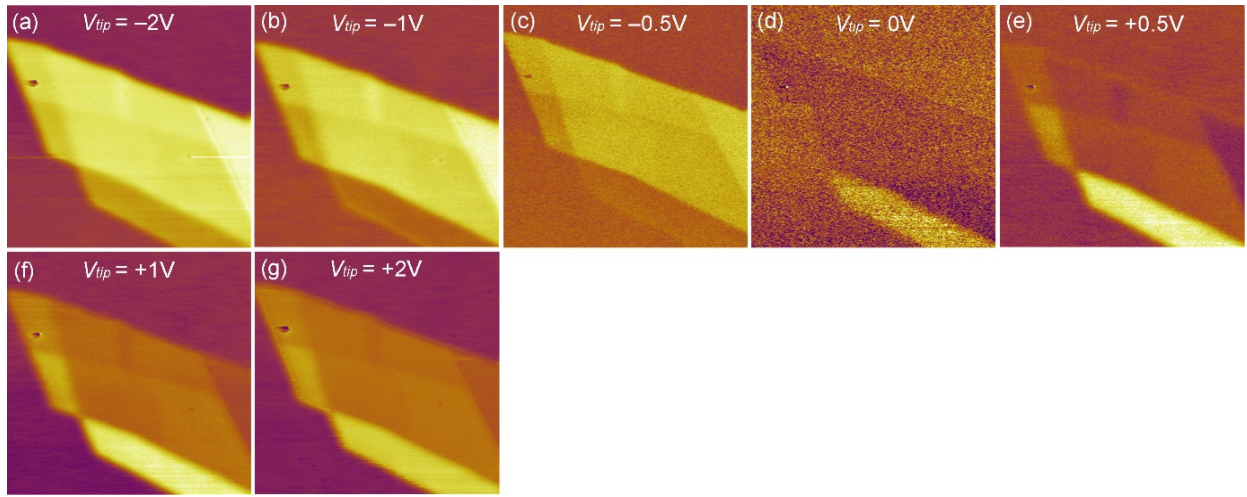

Figure 3. EFM images of the graphene nanosheets with $V_{tip}$ varying from –2 to +2 V ($z$ = 30 nm). The scale ranges: 3° for (a) and (g), 1° for (b) and (f), 0.4° for (c) and (e), 0.1° for (d).

The second test involves non-magnetic cantilever for "MFM" scans. A conductive but non-magnetic cantilever coated by Pt/Ti was used to scan graphene nanosheets following the scanning procedure that was in MFM, *i.e.* lift mode. Actually, such technique can be called electric force microscopy (EFM) which is sensitive to capacitive and electrostatic interactions but not magnetic interactions. Figure 3 shows that the EFM images are almost identical to the MFM images in Figure 2, except when $V_{tip} = 0$ (as in Figure 2d and 3d). The small difference is derived from the discrepancy of work function difference ($V_{WFD}$) between the cantilever tip and the graphene nanosheets due to different cantilever coating materials (Co/Cr *vs.* Pt/Ti), and possibly different work function and built-in charges of the MFM and EFM cantilevers ($V_Q$) (see Eq. 2). In fact, we found that magnetic and conductive cantilevers produced in the same batch could give rise to slightly different MFM and EFM phase images ($V_{tip} = 0$) of the same graphene nanosheets.[35] The reduced MFM phase signal with increased lift height (Figure 1e-i) can be also replicated by EFM,[35] because capacitive and electrostatic forces are also distance sensitive. It should be noted that similar EFM images from graphene have been reported before.[29,34,36] This test demonstrates that similar phase shift can be observed even when non-magnetic cantilever is used and re-confirms that strong capacitive and electrostatic interactions are present between conductive cantilever tip and the graphene on SiO₂/Si substrate. The above tests clearly show that the reported MFM signals mainly, if not completely, derives from non-magnetic interactions or artefacts.

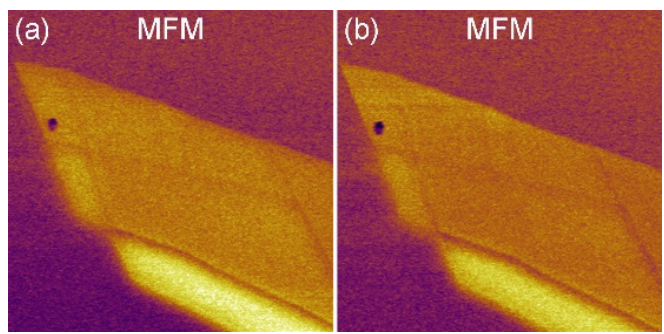

Figure 4. (a), (b) MFM phase images of the graphene nanosheets using magnetic cantilever tip with reversed direction of magnetic field ($z = 50$ nm). The scale range is 0.15°.

The third test is to see whether reverse of cantilever magnetism could have impact on the MFM signals. Reverse of tip magnetism is a common method to verify existence of ferromagnetism, because MFM phase shift should reverse (from attractive to repulsive, or vice versa) for ferromagnetic samples when the direction of tip magnetism is reversed; non-ferromagnetic samples, on the other hand, do not show MFM phase reverse. For instance, ferromagnetism of graphite was confirmed by this method.[3,5] Our tests show that the MFM phase images of the graphene nanosheets are very similar before and after the reverse of tip magnetism (Figure 4). In contrast, a floppy disk, as a control sample, did show expected MFM phase reverse after the magnetism of the same cantilever tip was reversed.[35] In addition, the effect of lift height on the MFM phase shift is identical for the reversed tip magnetism.[35] In fact, the reported MFM signals from whole pieces of graphene[21] are very different from the previously reported MFM signals from graphite which only shows magnetic signals at defects or edges, rather than from whole pieces of material.[3,5] These MFM results are therefore unlikely to be from magnetic interactions.

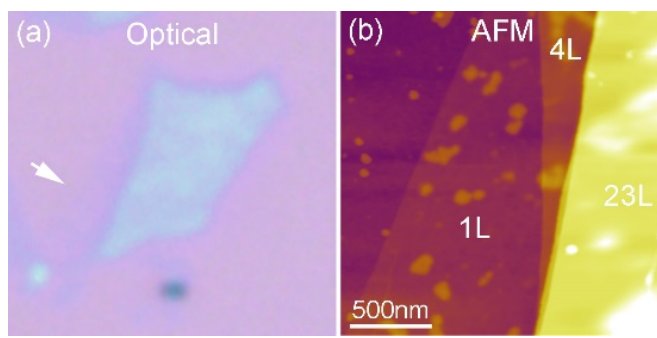

Figure 5. (a) Optical microscopy photo of the $MoS_2$ nanosheets on 90 nm $SiO_2$/Si substrate; (b) AFM image of the $MoS_2$ nanosheets of 1L, 4L and 23L thickness, and the scale range is 30 nm.

Similar MFM and EFM tests were conducted on the exfoliated $MoS_2$ nanosheets. Optical microscopy photo and AFM image of the $MoS_2$ nanosheets containing 1L (0.82 nm), 4L (2.98 nm) and 23L (15.9 nm) are shown in Figure 5a and b, respectively. Similar to the case of graphene, the MFM phase images of the $MoS_2$ nanosheets of different thicknesses show different contrast: the thicker $MoS_2$ (23L) exhibits more phase shift than the 1L and 4L nanosheets, consistent with the previous MFM study.[21] However, the MFM phase images of the $MoS_2$ nanosheets with $V_{tip}$ ranging from –2 to +2 V also show dramatic different phase shifts (Figure 6). The comparison of the vertexes of the fitted phase shift among the 1L, 4L and 23L $MoS_2$ nanosheets shows that the 1L $MoS_2$ (red) has a larger minimum of phase shift than the 4L (blue) and 23L (black) nanosheets (Figure 6i). This trend is opposite to that of graphene (Figure 2i), which can be attributed to $SiO_2$ substrate's negative (electron) doping for $MoS_2$ and positive (hole) doping for graphene.[29,30] This suggests that in addition to capacitive interactions, Coulombic (electrostatic) interactions between $MoS_2$ and the cantilever tip is strong, as supported by SKPM analysis which shows a strong charge effect (Figure 6j).[37] In addition, no MFM phase reverse was observed from the $MoS_2$ nanosheets under reversed tip magnetism either (Figure 7). So similar to the case of graphene, the MFM signals from $MoS_2$ nanosheets also have a significant non-magnetic contribution.

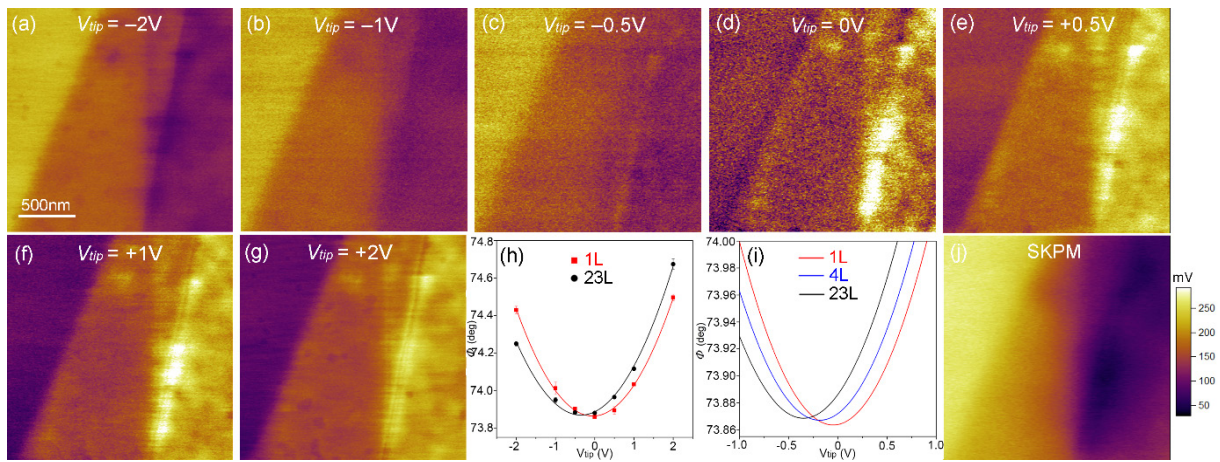

Figure 6. (a-g) MFM phase images of the $MoS_2$ nanosheets under different tip voltages ($V_{tip}$ = –2 to +2 V and $z$ = 50 nm). The scale ranges are 0.8° for (a) and (g), 0.4° for (b) and (f), 0.2° for (c) and (e), 0.1° for (d); (h) second degree polynomial fittings of the MFM phase values from the 1L and 23L $MoS_2$ nanosheets when $V_{tip}$ sweeping from –2 to +2 V; (i) fitted MFM

phase of 1L, 4L and 23L MoS$_2$ nanosheets as a function of $V_{tip}$. (j) SKPM image of the MoS$_2$ nanosheets using the same tip lift height.

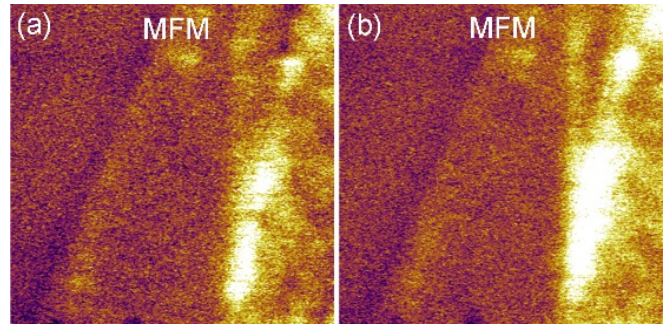

Figure 7. MFM phase image of the MoS$_2$ nanosheets with reversed tip magnetism. The scale range is 0.1°

## IV. CONCLUSION

Detailed investigations were conducted using MFM, EFM and SKPM to examine whether the observed MFM signals from exfoliated graphene and MoS$_2$ nanosheets can be correlated with their magnetism. Our MFM results which are similar to the previously reported confirm that graphene and MoS$_2$ nanosheets on 90 nm SiO2/Si substrate do show MFM phase contrast. However, it is found that the MFM response mainly, if not completely, comes from non-magnetic contributions including capacitive and electrostatic interactions. Therefore, the observed MFM signals from the exfoliated graphene and MoS$_2$ nanosheets does not relate to their magnetism, despite that these two materials could be magnetic and potentially show MFM signals.

## ACKNOWLEDGMENTS

Journal of Applied Physics 116, 213904 (2014). DOI: 10.1063/1.4903040

L.H. Li thanks Dr. Jiabao Yi from the University of New South Wales, Australia for valuable comments. Alfred Deakin Postdoctoral Research Fellowship and CRGS2013 grant are acknowledged for financial support.

# Supporting Information

# Electric Contributions to Magnetic Force Microscopy Response from Graphene and MoS$_2$ Nanosheets

Lu Hua Li[1]* and Ying Chen[1]

[1]Institute for Frontier Materials, Deakin University, Geelong Waurn Ponds Campus, Victoria 3216, Australia

*Corresponding author: luhua.li@deakin.edu.au

## 1. Effect of different cantilevers from the same batch

It is found that different cantilevers even from the same batch could give slightly different EFM results. Figure S1 shows the EFM images of the same piece of graphene nanosheets using two Pt/Ti coated conductive cantilevers produced in the same batch (same parameters, such as spring constant and coating material) with no voltage applied ($V_{tip}$ = 0V). It can be seen that the two EFM images are slightly different, which could be due to different trapped charges in the two cantilevers and their different tip structures, as discussed in Eq. 2.

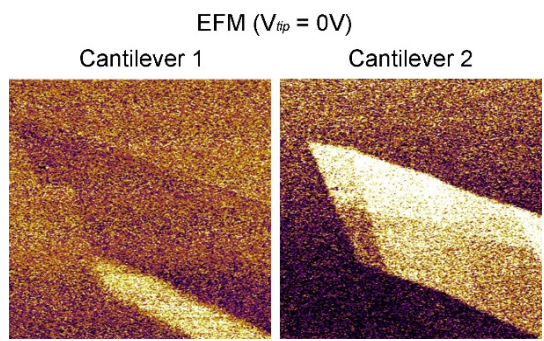

Figure S1. EFM images of the same piece of graphene nanosheets using two different cantilevers produced in the same batch. All the scanning conditions are identical.

## 2. Effect of lift height on EFM phase shift

The effect of lift height on the EFM phase images was shown in Figure 2, which also shows that the phase shift decreases with the increase of lift height, similar to that of the MFM phase images (Figure 1e-i).

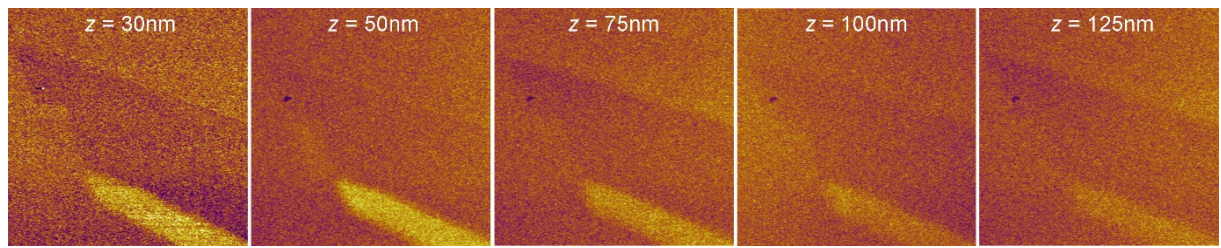

Figure S2. EFM phase images of the graphene nanosheets using different lift height (*z*). The scale range is 0.1° for all images.

## 3. MFM of a floppy disk before and after reverse of tip magnetism

MFM phase images were taken from a floppy disk with reversed direction of magnetic field of the magnetic cantilever tip. A particle on the surface of the floppy disk (* in Figure S3a) was used as a navigation mark to return to the same location after the reverse of tip magnetism. As expected, the phase shifts reversed with the reversed tip magnetism.

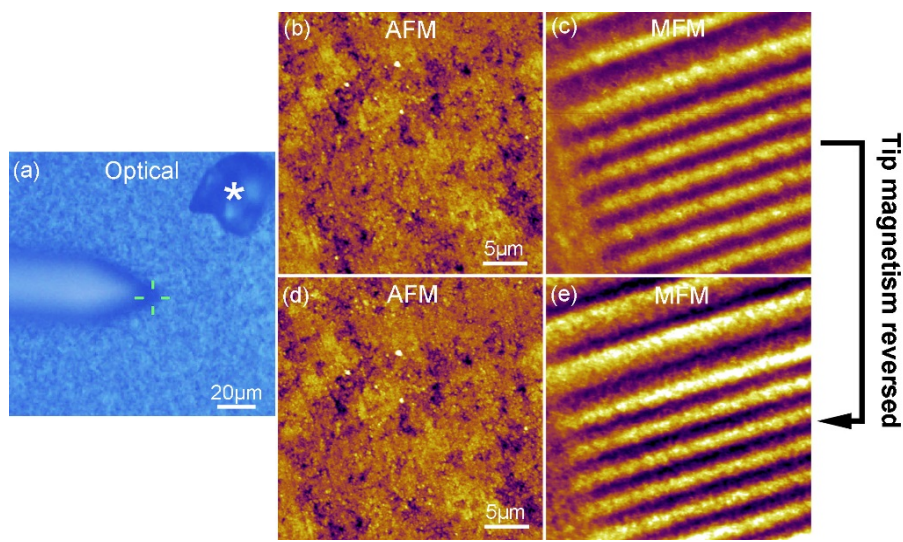

Figure S3. (a) Optical image of the surface of the floppy disk, where a particle (*) was used as a navigation mark so that the MFM images before and after the reverse of tip magnetism could be taken from the same location; (b), (c) AFM and MFM images before the reverse of tip magnetism; (c), (d) AFM and MFM images of the same location after the reverse of tip magnetism. The scale bars for the AFM topography and MFM phase images are 200 nm and 3.0°, respectively.

**4. Lift height effect on MFM phase images of the graphene with reversed tip magnetism**

The effect of lift height on the MFM phase images of the graphene is compared with reversed magnetism of the magnetic cantilever tip. Figure S4 shows that the MFM phase images at different lift heights ($z$ = 35 to 125 nm) are identical before and after the tip magnetism was reversed.

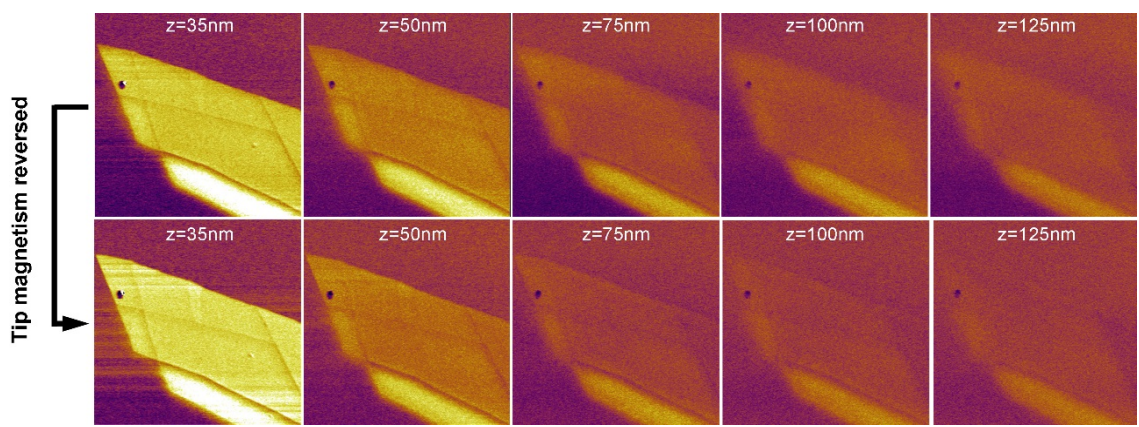

Figure S4. MFM phase images of the graphene nanosheets under the initial direction of magnetic field of the magnetic cantilever tip (upper row) and MFM images under the reversed tip magnetism (lower row).